\documentclass[aps,prd,twocolumn,groupedaddress,nofootinbib]{revtex4-2}
\usepackage{graphicx}
\usepackage{xcolor}
\usepackage{amsmath}
\usepackage{ulem}
\usepackage{hyperref}
\usepackage{aas_macros}
\begin{document}
%\preprint{}

\title{Implications of photon-ALP oscillations in the extragalactic neutrino source TXS 0506+056 at sub-PeV energies}


\author{Bhanu Prakash Pant}
\email{pant.3@iitj.ac.in}
%\thanks{}
\author{Sunanda}
\author{Reetanjali Moharana}
\email{reetanjali@iitj.ac.in}
%\email{sunanda@iitj.ac.in}
\author{Sarathykannan S.}
%\email{sarathy.1@iitj.ac.in}
\affiliation{Department of Physics, Indian Institute of Technology Jodhpur, Karwar 342037, India.}
%\thanks{}

\date{\today}

\begin{abstract}
Photon-axion-like particle (ALP) oscillations result in the survival of gamma rays from distant sources above TeV energies. Studies of events observed by CAST, Fermi-LAT, and IACT have constrained the ALP parameters. We investigate the effect of photon-ALP oscillations on the gamma-ray spectra of the first extragalactic neutrino source, TXS 0506+056, for observations by Fermi-LAT and MAGIC around the IC170922-A alert. We obtain a constraint on the ALP coupling parameter $g_{a\gamma} < 5 \times 10^{-11}$ GeV$^{-1}$ with 95\% C.L. when focusing on the ALP mass range 0.1 neV $\le$ $m_a$ $\le$ 1000 neV. Importantly, we study the implications of ALP-$\gamma$ oscillations on the counterpart $\gamma$ rays of the sub-PeV neutrinos observed from TXS 0506+056. We also show the diffuse $\gamma$-ray fluxes and observabilities from flat-spectrum radio quasars, high-synchrotron peaked sources, and low-intermediate-synchrotron peaked sources, assuming similar gamma-ray emissions as that from TXS 0506+056. %Our  We studied the quiescent and the two VHE flaring phases 
%of the blazar to set constraints on the ALP parameter space. 
%and additionally, the 2014 neutrino flare phase for the source to study the ALP effect at sub-PeV energies.
\end{abstract}

%\keywords{}

\maketitle
\section{\label{sec:intro}Introduction}
Axion-like particles (ALPs) are pseudoscalar (spin-0) bosons with very light mass and are potential candidates for dark matter \cite{preskill1983cosmology,khlopov1999nonlinear}. The axions are also proposed to solve the $CP$ problem in QCD \cite{peccei1977constraints,sikivie2010dark}. %and weak coupling to standard model (SM) particles. They are considered as a potential dark matter (DM) candidate and thus may account for a significant fraction of cold dark matter in the Universe \cite{preskill1983cosmology,abbott1983cosmological,dine1983not,khlopov1999nonlinear,sikivie2010dark,marsh2016axion}, also an extension of physics beyond the standard model (BSM) \cite{svrcek2006axions,arvanitaki2010string}, similar to QCD axions that are proposed originally by Peccei and Quinn to solve the strong CP problem in QCD \cite{peccei1977constraints,peccei1977cp}.  
In the presence of an external magnetic field, ALPs can couple to photons via two coupling vertices which leads to the photon-ALP oscillation. %effect in which the free photon and ALP state may convert into each other. 
In astrophysical environments, photon-ALP conversion drastically reduces the absorption of very-high-energy (VHE) photons by extragalactic background light (EBL) and the cosmic microwave background (CMB) through pair production above 100 GeV \cite{de2011relevance,reesman2014probing,tavecchio2015photons, galanti2019blazar, li2022searching, li2022probing}. 

This increased transparency can modulate and enhance the observed $\gamma$-ray spectra of the TeV photons originating from higher-redshift sources using observations of $\gamma$-ray spectra from VHE sources \cite{hooper2007detecting,simet2008milky,tavecchio2012evidence,meyer2013first, abramowski2013constraints, ajello2016search, berenji2016constraints, kohri2017axion,  meyer2017fermi, galanti2018extragalactic, zhang2018new, liang2019constraints,long2020testing,libanov2020impact, bi2021axion, guo2021implications, gautham2021reconciling, li2021limits} to set stringent constraints on the ALP mass, $m_a$, and coupling constant $g_{a\gamma}$. Interestingly, the recent observations of nearly 18-TeV photons by the Large High Altitude Air Shower Observatory (LHAASO) with the kilometer-square area (KM2A) \cite{lhaasogcn} and an astonishing 251-TeV photon by Carpet-2 \cite{carpetgcn} from the long gamma-ray burst GRB 221009A at redshift 0.1505 has motivated the community to understand the survival of photons at this energy through ALP-photon oscillation \cite{galanti2022grb, baktash2022grb}. We note that the above-mentioned 251-TeV photon observed by Carpet-2 also has the candidate sources LHAASO J1929+1745 and 3HWC J1928+178, as reported in Ref. \cite{Fraija}. Most of these studies focused on photons at energies observed by the Imaging Atmospheric (or Air) Cherenkov Telescope (IACT). Observations from the axion flux of the Sun have also been studied by the CERN Axion Solar Telescope (CAST), giving the most stringent constraint on the ALP parameters,  $m_a < 0.01$ eV,  $g_{a\gamma} <$ 6.6$\times$10$^{-11}$ GeV$^{-1}$ \cite{zioutas1999decommissioned, 2017NatPh..13..584A}. %and using observations of $\gamma$-ray spectra from VHE sources \cite{de2007evidence,hooper2007detecting,simet2008milky,belikov2011no,de2011relevance,tavecchio2012evidence,meyer2013first, abramowski2013constraints, reesman2014probing, meyer2014detecting, tavecchio2015photons, ajello2016search, berenji2016constraints, meyer2017fermi, galanti2018extragalactic, zhang2018new, galanti2019blazar,liang2019constraints,long2020testing,libanov2020impact, bi2021axion, guo2021implications, gautham2021reconciling, li2021limits} to set stringent constraints on ALP mass and coupling constant.}

The effect of ALP-photon oscillation at sub-PeV energies and higher has recently been explored with Galactic diffuse gamma rays using High Altitude Water Cherenkov (HAWC), Tibet AS-$\gamma$, and LHAASO events, resulting in a limit of $m_a < 2 \times 10^{-7}$ eV,  $g_{a\gamma} < 2.1 \times 10 ^{-11}$ GeV$^{-1}$ with 95\% confidence limit (C.L.) \cite{eckner2022}, and excluding $g_{a\gamma} > 3.9 - 7.8 \times 10 ^{-11}$ GeV$^{-1}$ for $m_a < 4 \times 10^{-7}$ eV
 \cite{mastrotaro2022} at 95\% C.L., respectively. The intrinsic photon flux would be different at these energies than at lower energies due to the addition of  hadronic channels. With the observations of sub-PeV neutrinos, we may understand the energetics of the source.
  
In this work, we investigate the implications of photon-ALP oscillation for the first ever non-Galactic sub-PeV neutrino source \cite{10.1093/mnras/sty1852} TXS 0506+056 situated at a redshift $z_{0}=0.3365$ \cite{paiano2018redshift}. It was first discovered as a radio source \cite{lawrence19835} and later as high-energy gamma radiation with space missions, like the Energetic Gamma Ray Experiment Telescope and Fermi-Large Area Telescope (\textit{Fermi}-LAT)\cite{halpern2003redshifts, lamb1997point, abdo2010first}. %This is the first-ever non-stellar a source associated with sub-PeV neutrinos of astrophysical origin \cite{10.1093/mnras/sty1852}. 
 On September 22, 2017 (IC170922-A), the IceCube Neutrino Observatory detected a very-high-energy $\sim$ 290-TeV muon neutrino coinciding with the direction of a flaring state of TXS 0506+056 \cite{icecube2018multimessenger}. Soon, follow-up observations were performed in various energy bands by \textit{Fermi}-LAT ($\gamma$ rays) \cite{tanaka2017astronomer}, the Nuclear Spectroscopic Telescope Array (X rays) \cite{2017ATel10845....1F}, and Swift (X rays, UV, optical) \cite{2017ATel10792....1E}, and VHE $\gamma$-ray observations were made by the Major Atmospheric Gamma Imaging Cherenkov Telescopes (MAGIC) \cite{2017ATel10817....1M}, High Energy Stereoscopic System \cite{de2017hess}, HAWC \cite{martinez2017hawc} and Very Energetic Radiation Imaging Telescope Array System \cite{mukherjee2017veritas}. Notably, prior to the IC170922-A alert, this source was also observed with a neutrino flare, making it a sub-PeV neutrino source with a significance of 3.5$\sigma$ \cite{icecube2018neutrino}. However, there was no significant flaring in MeV-GeV gamma rays during this epoch. A hadronic-originated photon counterpart will contribute to the intrinsic flux at sub-PeV for such neutrino sources. Hence, TXS 0506+056 is the candidate source to study the ALP-photon oscillation at several-TeV to sub-PeV energies.
 
 This paper is organized as follows. In Sec. \ref{sec:alpeffect}, we describe the propagation of photon-ALP beam in an external magnetic field. In Sec. \ref{sec:magmodel} we describe the various magnetic field models used for the analysis. In Sec. \ref{sec:method} we discuss the methodology used for data fitting and predicting the expected $\gamma$-ray flux. In Sec. \ref{sec:datasel} we describe the significance of the ALP effect in TXS 0506+056 and its \textit{Fermi}-LAT analysis. In Sec. \ref{sec:resdis} we discuss the results for the ALP-$\gamma$ oscillations. This section also includes the implication of this oscillation for the diffuse gamma-ray flux from TXS 0506+056-like sources. Here we calculate the diffuse gamma-ray flux from sources, flat-spectrum radio quasars(FSRQs), high-synchrotron peaked (HSP) sources,
and low-intermediate-synchrotron peaked (LISP) sources, and their future observability. 

\section{\label{sec:alpeffect}Photon-ALP Oscillation and Propagation in Magnetic Fields}

The minimal interaction coupling $g_{a\gamma}$ between photons of energy $E'_\gamma$ and ALPs in the presence of an external magnetic field \textbf{B} and electric field \textbf{E} has been proposed in the literature \cite{raffelt1988mixing, de2011relevance}.

A polarized, monoenergetic photon beam propagating along the $\hat{\textbf{z}}$ direction in a cold plasma medium with a homogeneous \textbf{B} field along the $\hat{\textbf{y}}$ axis, has the equation of motion,
%In the presence of an external magnetic field, the interaction between photon and ALP under the minimal photon-ALP coupling can be described as,
%\begin{equation}
 %   \mathcal{L}_{int} = \frac{-1}{4}g_{a\gamma} \,a\,F_{\mu \nu} \tilde{F}^{\mu \nu}=g_{a\gamma}\,a\,\textbf{E} \cdot \textbf{B},
%\end{equation}
%where $g_{a\gamma}$ is the coupling between photons and ALP, $F_{\mu \nu}$ is the electromagnetic field tensor, $\tilde{F}^{\mu \nu}$ is the dual tensor, \textbf{E} is the electric field, and \textbf{B} is the magnetic field, respectively. 

%We consider initially polarized, monoenergetic beam of photons with energy $E^{'}_{\gamma}$ propagating along the $\hat{\textbf{z}}$ direction. The equation of motion in the homogeneous external magnetic field \textbf{B} covered over a medium of cold plasma is \cite{raffelt1988mixing, de2011relevance},   
%Considering the cold plasma medium through which the beam is propagating to be filled with homogeneous external magnetic field \textbf{B}, and in the limit $E \gg m_{a}$, the equation of motion is given by \cite{raffelt1988mixing, de2011relevance}:
\begin{equation}
\left(i\frac{d}{dz}+E^{'}_{\gamma}+\mathcal{M}_{0}\right) \psi(z) = 0 \,\Big|_{E'_\gamma \gg m_{a}}\, ,
\end{equation}
where $\mathcal{M}_{0}$ represents the photon-ALP mixing matrix and $\psi(z) = \begin{pmatrix}
A_1(z) & A_2(z) & a(z)
\end{pmatrix}^T$ denotes the state function. Here, $A_1(z)$ and $A_2(z)$ are the photon amplitudes with linear polarizations along the x and y axis, respectively, whereas $a(z)$ is the amplitude associated with the ALP state.

%We consider the homogeneous external magnetic field \textbf{B} along the $\hat{\textbf{y}}$ axis, so that $B_{x}=0$. 
Assuming weak magnetic fields and $E^{'}_{\gamma}$ at VHE, the QED vacuum polarization and Faraday rotation can be neglected, and the mixing matrix becomes
%restricting to deal with weak magnetic fields only, we can neglect the contribution from the QED vacuum polarization  In addition, we can neglect the effect of Faraday rotation since we are considering the energy $E$ in the  VHE $\gamma$-rays regime. This leads to the simplification of the form of mixing matrix $\mathcal{M}_{0}$,   
\begin{equation}
\mathcal{M}_{0} = 
\begin{pmatrix}
\Delta^{xx} & 0 & 0\\
0 & \Delta^{yy} & \Delta^{y}_{a\gamma} \\
0 & \Delta^{y}_{a\gamma} & \Delta^{zz}_{a}
\end{pmatrix}
\, ,
\end{equation}
where $\Delta^{xx} = \Delta^{yy} = -\omega^{2}_{pl}/2E $, $\Delta^{zz}_{a} = -m^{2}_{a}/2E$, and $\Delta^{y}_{a\gamma} = g_{a\gamma\gamma}B_{y}/2$. Here, $\omega^{2}_{pl}$ is the plasma frequency resulting from the charge-screening effect.
%which results from the effective photon mass arises due to the charge screening effect as the beam is propagating through the cold plasma.
The propagation region of the photon-ALP beam is divided into N subregions. In each region, the probability for photon survival is calculated. The initial beam state is 
\begin{equation}
    \rho(0) = \frac{1}{2} \,diag.(1, 1, 0).
\end{equation}
%The transport matrix $T(s)$ of the photon-ALP beam can be written by splitting the whole propagation length into N sub-regions,
%\begin{equation}
 %   T(s) = T(s_{N}) \times T(s_{N-1}) \times ... \times T(s_{1}),
%\end{equation}
%where $T(s_{1})$, $T(s_{2})$,..., $T(s_{N})$, are the transport matrices for each of the sub-regions.

%Assuming the initial beam state $\rho(0)$ to be,
%\begin{equation}
%    \rho(0) = \frac{1}{2} \,diag.(1, 1, 0).
%\end{equation}

The final photon survival probability can be written as:
\begin{equation} 
\label{galp-prob}
    P_{\gamma\gamma} = Tr\left[(\rho_{11}+\rho_{22})T(s)\rho(0)T^{\dagger}(s))\right],
\end{equation}
where $\rho_{11}$ $=$ diag(1, 0, 0) and $\rho_{22}$ $=$ diag(0, 1, 0) denote the polarization along the x and y axis, respectively, and $T(s) = T(s_3)_{Gal} \times T(s_2)_{Ext} \times T(s_1)_{Source}$ is the whole propagation transfer matrix. Here, the subregions are the source, the extragalactic medium, and the Milky Way region.%, to obtain the transfer matrix, and consequently the final survival probability of the photon-ALP system. 

\section{\label{sec:magmodel}Magnetic field models}
%In order to predict the expected $\gamma$-ray spectrum at Earth, it is crucial to know the magnetic field environments through which $\gamma$-rays from the source are expected to propagate. 
In this section, we give a brief overview of the magnetic field models used in our calculation. 
\subsection{Blazar jet region}
We consider the photon-ALP oscillation at the source in the presence of blazar jet magnetic field (BJMF). The magnetic field in the jet region can be modeled with poloidal (along the jet axis, $B \propto r^{-2}$) and toroidal (transverse to the jet axis, $B \propto r^{-1}$) components. At distances large enough from the central black hole, the toroidal component dominates over the poloidal component and thus the latter can be neglected. We adopt the toroidal magnetic field strength $B^{jet}(r)$ given by \cite{begelman1984theory,ghisellini2009canonical}:
\begin{equation}
    B^{jet}(r) = B^{jet}_{0}\left(\frac{r}{r_{VHE}}\right)^{\eta}, \label{eq:bfieldjet}
\end{equation}
where $B^{jet}_{0}$ is the magnetic field strength at the core and $r_{VHE}$ is the distance between the VHE $\gamma$-ray-emitting region and the central black hole. We assume $r_{VHE} \sim R_{VHE}/\theta_{j}$, where $R_{VHE}$ is the blob radius of the VHE emitting region and $\theta_{j}$ is the angle between the jet axis and the line of sight. 

Assuming equipartition between the magnetic field and particle energies, the electron density profile $n_{el}(r)$ can be modeled as a power law given by \cite{o2009magnetic}:
\begin{equation}
    n_{el}(r) = n_{0}\left(\frac{r}{r_{VHE}}\right)^{\xi}, \label{eq:eldensjet}
\end{equation}
where $n_{0}$ is the electron density at $r_{VHE}$. In Ref. \cite{davies2022relevance}, a more realistic model was provided that takes into account the fact that the electron distribution is nonthermal in a relativistic active galactic nuclei Jet.

%The above equations, namely Eq. (\ref{eq:bfieldjet}) and Eq. (\ref{eq:eldensjet}), hold in the comoving jet frame with 
The photon energy $E^{'}_{\gamma}$ in the jet frame is related to the lab-frame energy $E_{\gamma}$ by $E^{'}_{\gamma} = E_{\gamma}/\delta_{D}$, where $\delta_{D} = \left[\Gamma_{L}(1-\beta_{j}^{2}cos\theta_{j})\right]^{-1}$ is the Doppler factor with $\Gamma_{L}$ and $\beta_{j}$ being the bulk Lorentz and beta factor, respectively. We assume that at $r>1$ kpc the BJMF strength is negligible. Further details of the BJMF model can be found in Refs. \cite{tavecchio2015photons, galanti2019blazar}.

\subsection{Intercluster magnetic fields}
%If the blazar is located in a cluster rich environment, the turbulent magnetic field of $\sim$ $\mathcal{O}$(1)$\mu$G \cite{carilli2002cluster,govoni2004magnetic,subramanian2006evolving} induces a significant conversion between photons and ALPs \cite{meyer2014detecting, wouters2012irregularity}. 

The intercluster medium magnetic field (ICMF), $B^{ICMF}$ can be modeled as,
\begin{equation}
    B^{ICMF}(r) = B_{0}^{ICMF} \left(\frac{n_{el}(r)}{n_{el}(r_{0})}\right)^{\eta},
\end{equation}
with $0.5\le\eta\le1.0$, where $B_{0}^{ICMF}$ and $n_{el}(r_{0})$ are the magnetic field strength and electron density at the cluster center, respectively. The electron density distribution $n_{el}(r)$  at a distance $r$ from the cluster center is
\begin{equation}
    n_{el}(r) = n_{0}^{ICMF} \left(1 + \frac{r}{r_{core}}\right)^{\zeta},
\end{equation}
with $r_{core}$ is the core radius and $\zeta=-1$. The typical order of the electron density and the core radius is $\sim \mathcal{O}(10^{-3})$ cm$^{-3}$ and $\sim \mathcal{O}(100)$ kpc, respectively.

In a cluster-rich environment, the turbulent magnetic field is of order $\sim$ $\mathcal{O}$(1)$\mu$G \cite{carilli2002cluster,govoni2004magnetic,subramanian2006evolving}. Such a cluster can have a significant effect on the conversion between photons and ALPs \cite{meyer2014detecting, wouters2012irregularity}. Since there is no evidence that TXS 0506+056 is located in a cluster-rich environment, we do not consider the photon-ALP oscillations in the ICMF model.

%In case the blazar is located in a cluster rich environment, the turbulent magnetic field can be of $\sim$ $\mathcal{O}$(1)$\mu$G \cite{carilli2002cluster,govoni2004magnetic,subramanian2006evolving}. This induces a significant conversion between photons and ALPs \cite{meyer2014detecting, wouters2012irregularity}. 

\subsection{Extragalactic magnetic fields}
The actual strength of the extragalactic magnetic field on the cosmological scale $\sim \mathcal{O}$(1) Mpc, is still unknown, but the currently accepted limit is $\sim\mathcal{O}$(1) nG \cite{ade2016planck, pshirkov2016new}. In this work, we neglect the effect due to the magnetic field (See Ref. \cite{ galanti2018behavior} for possible effects of the magnetic field) and consider only the absorption effect due to EBL. 

%The $\gamma$-rays propagating across the intergalactic medium undergoes pair production process  $\gamma + \gamma_{BG} \rightarrow e^{+} + e^{-} $ with Cosmic Microwave Background (CMB) radiation. 
The optical depth of EBL attenuation can be written as \cite{belikov2011no}
\begin{eqnarray}
    \tau (E_{\gamma},z_{0}) = c\,\int_{0}^{z_{0}} \frac{dz}{(1+z)H(z)} \, \int_{E_{th}}^{\infty} dE^{BG}_{\gamma} \, \frac{dn(z)}{dE^{BG}_{\gamma}} \, \nonumber \\ \times \, \tilde{\sigma}(E_{\gamma}, E^{BG}_{\gamma})\,,
\end{eqnarray}
with
\begin{eqnarray}
    \tilde{\sigma}(E_{\gamma}, E^{BG}_{\gamma}) = \int_{-1}^{1-\frac{2(m_ec^2)^2}{E^{BG}_{\gamma}E_{\gamma}}} \, dcos\theta \, \frac{(1-cos\theta)}{2} \, \nonumber \\ \times \, \sigma_{\gamma\gamma}(E_{\gamma}, E^{BG}_{\gamma},\theta) \, ,
\end{eqnarray}
and
%\begin{equation}
 $ E_{th} = 2 \cdot (m_{e}c^{2})^{2}/ (E_{\gamma}(1-cos\theta)), $
%\end{equation}
where $z_{0}$ is the redshift of the source, $H(z)$ is the rate of Hubble expansion, $E_{th}$ is the threshold energy for pair production, $dn(z)/dE^{BG}_{\gamma}$ is the proper number density of the EBL, $\sigma_{\gamma\gamma}(E_{\gamma}, z, E^{BG}_{\gamma})$ is the pair-production cross section, $\theta$ is the angle between the projectile and target photons, $E_{\gamma}$ is the projectile photon energy, and $E^{BG}_{\gamma}$ is the target background photon energy. Several EBL models have been proposed in the literature \cite{franceschini2008extragalactic, kneiske2010lower, dominguez2011extragalactic, gilmore2012semi, inoue2013extragalactic, franceschini2017extragalactic, saldana2021observational}, and we consider the model from Ref. \cite{dominguez2011axion} in this work. 

\subsection{Milky Way region}
Finally, we consider the effect of photon-ALP oscillation in the presence of a Galactic magnetic field (GMF). This effect can have both a large-scale regular component and a small-scale random component. Due to the fact that the coherence length is smaller than the oscillation length, we neglect the random component \cite{meyer2014detecting} in our analysis and consider only the regular GMF component model given in \cite{jansson2012new}; the latest model can be found in Refs. \cite{Jansson_2012, adam2016planck}.

\section{\label{sec:method}Methodology}
%The intrinsic energy spectrum of the $\gamma$-ray blazars can be described by simple functions which may assume three to five parameters. 
The photon beam of the blazar can be considered as the intrinsic photons generated by the accelerated leptons or hadrons. In general, one can consider the intrinsic spectrum to follow the superexponential cutoff power law (SEPWL) to fit the observed data points under the null hypothesis,
\begin{equation}
    \Phi_{int}(E) = N_{0} \left(\frac{E}{E_{0}}\right)^{-\alpha} exp\left[-\left(\frac{E}{E_{cutoff}}\right)^\beta\right],
\end{equation}
where $E_{0}$ is taken to be 1 GeV, and $N_{0}$, $\alpha$, $E_{cutoff}$, and $\beta$ are treated as free parameters. The best-fit parameters are given in Table \ref{table:bestfitparams}. It is to be noted that we also tested other forms of the intrinsic spectrum and found that the SEPWL gives the smallest value of the best-fit $\chi^{2}$ under the null hypothesis.

\begin{table}
\begin{ruledtabular}
\caption{\label{table:bestfitparams}Summary of the best-fit spectral parameters for all phases.}
\begin{tabular}{c c c c c c}
 Phase & $N_0$ (x10$^{-11}$) & $\alpha$ & $\beta$ & $E_{cutoff}$ \\ 
  & [MeV$^{-1}$cm$^{-2}$s$^{-1}$] & & & [GeV]\\
 \hline
 Neutrino Flare 2014 & 0.65 & 1.79 & 0.1 & 71.01\\
VHE Flare 1 & 4.04 & 1.99 & 0.54 & 66.27\\
VHE Quiescent & 1.59 & 1.94 & 0.63 & 58.37\\
\end{tabular}
\end{ruledtabular}
\end{table}

We use the open-source PYTHON-based package \texttt{gammaALPs}\footnote{\url{https://gammaalps.readthedocs.io/en/latest/index.html}} \cite{Meyer:202199} to compute the photon-ALP oscillation probability $P^{ALP}_{\gamma\gamma}$.% in different astrophysical environments as described in Section \ref{sec:magmodel}.
\begin{figure*}
    \centering
    \includegraphics[width=0.85\textwidth]{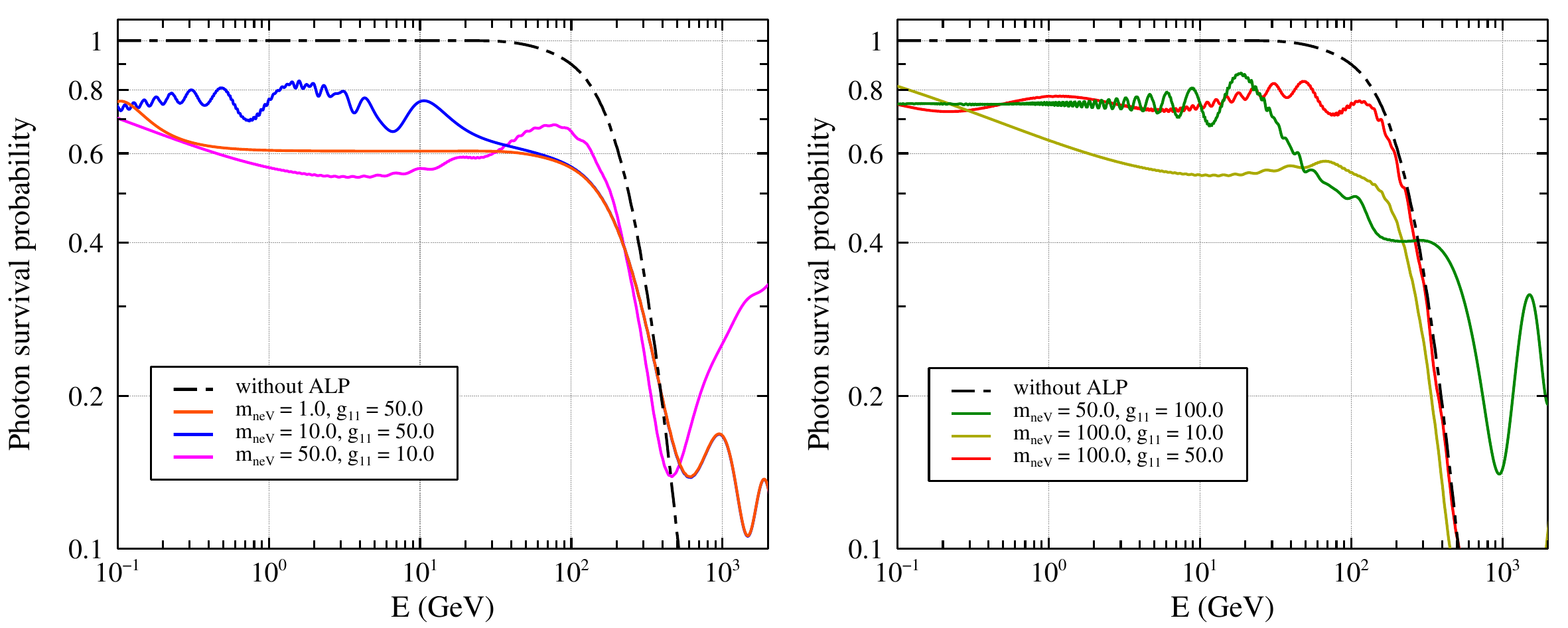}
    \caption{\label{fig:probalp_q}Final photon survival probability for the TXS 0506+056 blazar for six typical sets of parameters. The dot-dashed and solid lines represent the photon survival probability without and with the ALP effect, respectively. The notations used to represent the parameters are m$_{neV}$ $\equiv$ m$_{a}$/1 neV and g$_{11}$ $\equiv$ g$_{a\gamma}$/10$^{-11}$ GeV$^{-1}$.}
\end{figure*}

The modulated $\gamma$-ray spectrum obtained after the photon-ALP oscillation is given by  %\begin{equation}
   $ \Phi_{w ALP}(E) = P^{ALP}_{\gamma\gamma} \Phi_{int}(E)$.
%\end{equation}
In order to get the expected $\gamma$-ray spectrum observed in the detector, we must consider the energy resolution of the detector. One can assume the energy dispersion function $D(E_{t}, E^i_{\gamma}, E^j_{\gamma})$ to be a Gaussian with the variance being the energy resolution. Then the expected flux between energy bins $E^i_{\gamma}$ and $E^j_{\gamma}$ is given by \cite{guo2021implications}   
\begin{equation}
    \Phi^{exp}(E_\gamma) = \frac{\int_{0}^{\infty} D(E_{t}, E^i_{\gamma}, E^j_{\gamma}) \, \Phi_{wALP}(E_{t}) \, dE_{t}}{E^i_{\gamma} - E^j_{\gamma}},
\end{equation}
where $E_{t}$ is the true energy of $\gamma$ rays. The energy resolution of \textit{Fermi}-LAT and MAGIC are taken as $15\%$ \footnote{\url{https://fermi.gsfc.nasa.gov/ssc/data/analysis/documentation/Cicerone/Cicerone_Introduction/LAT_overview.html}} and $16\%$ \cite{aleksic2016major}, respectively.

One can obtain the best-fit photon-ALP oscillation probability by fitting the observed flux with the expected flux. We define $\chi^{2}$ as,
\begin{equation}
    \chi^{2} = \sum_{i}^{N_{bins}} \left(\frac{\Phi^{exp}_{i}(E_\gamma) - \Phi^{obs}_{i}(E_\gamma)}{\sigma_{i}}\right)^{2} \, ,
\end{equation}
where $\Phi^{obs}_{i}$ is the observed $\gamma$-ray flux, with $\sigma_{i}$ being the corresponding uncertainty. The best-fit $m_{a}$ and $g_{a\gamma}$ are calculated by minimizing the $\chi^{2}$. 
\section{\label{sec:datasel}Significance of ALP effect in TXS 0506+056}

The first extragalactic TeV-PeV neutrino source TXS 0506+056 is the best candidate to study high-energy radiation. The follow-up observations of IC170922-A showed that the source has multiwavelength emissions. Additionally, the source showed an excess of $13 \pm 5$ in the time window of 158 days, from September 2014 (MJD 56937.81) to March 2015 (MJD 57096.21), known as the neutrino flare phase \cite{icecube2018neutrino}. Recently, $\sim$ 41-hr and $\sim 74$-hr observations of VHE ( $> 90$ GeV) events by MAGIC between September 2017 to December 2018 showed three flaring activities \cite{ansoldi2018blazar, acciari2022investigating}. Hence, the blazar TXS 0506+056 would be the best probe to study the photon-ALP interaction in the MeV to the sub-PeV range. We study the VHE activity phases as well as the neutrino-flare phase of the source. 

\subsection{\textit{Fermi}-LAT analysis of TXS 0506+056}

We analyze the \textit{Fermi}-LAT data for TXS 0506+056 in three phases: 
\begin{itemize}
    \item \textbf{Neutrino Flare :} September 2014  (MJD 56937.81) to March 2015  (MJD 57096.21).
    \item \textbf{VHE Flare 1 :} 04 September, 2017 (MJD 58000) to 03 November, 2017 (MJD 58060).
    \item \textbf{VHE Quiescent :} 04 November, 2017 (MJD 58061) to 10 October, 2018 (MJD 58422).
\end{itemize}

%Another episode of flaring activity occurs from 2017 September 04 (MJD 58000) to November 03 (MJD 58060). Then it undergoes low activity (quiescent) phase, from 2017 November 04 (MJD 58061) to 2018 October 10 (MJD 58422). Then, it again enters into a flaring state from 2018 September 01 (MJD 58423) to 2019 January 02 (MJD 58485).

We select the above-mentioned time period of \textit{Fermi}-LAT (Pass 8) processed data from the Fermi Science Data Center \footnote{\label{foot:fermi}\url{https://fermi.gsfc.nasa.gov/ssc/data/access/}}. We choose the spacecraft files and the instrument response functions (IRFs) of \texttt{P8R3\_SOURCE\_V2} to match the extracted data. We select the \texttt{SOURCE} event class (evclass=128 and evtype=3) in the energy range of 100 MeV to 300 GeV. The region of interest is selected to be 2$^{\circ}$ centered on the target source. We consider all of the 4FGL sources around 10$^{\circ}$ of TXS 0506+056 as background sources together with preprocessed templates of Galactic diffuse emission, \texttt{gll\_iem\_v08.fits} , and the extragalactic isotropic diffuse emission, \texttt{iso\_P8R3\_SOURCE\_V2\_v2.fits}.

The likelihood analysis, spectral energy distribution, and light curve are obtained by using the PYTHON-based Fermipy package \footnote{\url{https://fermipy.readthedocs.io/en/latest/index.html}} \cite{2017ICRC...35..824W}.

\section{\label{sec:resdis}Results and Discussions}
We divide this section into two important parts: results obtained by calculating the oscillation probability, and discussions on the possible gamma rays at sub-PeV energies resulting from this oscillation. 

\subsection{\label{sec:resultsA}Constraints on ALP parameter space}
We calculate the photon-ALP oscillation probability for the three phases of TXS0506+056 mentioned above following Sec. \ref{sec:method}.
%\subsection{\label{sec:const}Constraints on ALP Parameter Space}

\subsubsection{\label{subsec:resultsA1}VHE activity around IC170922-A}
In this section, we discuss the results obtained for VHE Flare 1, and VHE Quiescent associated with the IC170922-A event using the \textit{Fermi}-LAT and MAGIC events. The VHE events observed by MAGIC are collected from Ref. \cite{ansoldi2018blazar, acciari2022investigating}. We use the modeling parameters for these phases from \cite{acciari2022investigating}. The parameters used in the  BJMF model in the \texttt{gammaALPs} package are listed in Table \ref{table:source_params}. Note that for this study we do not consider the ICMF model. 
\begin{table}[ht!]
\begin{ruledtabular}
\caption{\label{table:source_params}Summary of the BJMF model parameters in the quiescent and flaring states of TXS0506+056 taken from Ref. \cite{acciari2022investigating}.}
\begin{tabular}{c c c}
 \textbf{Parameter name} & \textbf{Quiescent} & \textbf{Flaring}  \\ 
 \hline
 R.A.(J2000) & 05 09 25 (hh mm ss) & "\\
 Dec.(J2000) & +05 42 09 (dd mm ss) & "\\ 
 z & 0.337 & " \\
$\theta_{view}$ [deg] & 0.8 & " \\ 
 $\delta$ & 40 & " \\
 $\Gamma$ & 22 & "  \\
 $B^{Jet}_{0}$ [G] & 1 & "  \\
 $n^{Jet}_{0}$ [cm$^{-3}$] & $10^{3}$ & 520 \\
 $r_{VHE}$ [pc] & 10 & "\\
 $R^{'}_{blob}$ [$10^{16}$ cm] & 1.1 & "\\
 $\eta$ & -1 & " \\
 $\xi$ & -2 & " \\
 $\gamma_{e,min}$ & 800 & " \\
 $\gamma_{e,max}$ & $10^{4}$ & $2 \times 10^{4}$ \\
\end{tabular}
\end{ruledtabular}
\end{table}

Figure \ref{fig:probalp_q} shows the photon survival probability due to EBL/CMB and photon-ALP oscillations for several typical sets of $m_a$ and $g_{a\gamma}$.
%In Table \ref{table:source_params}, we list the best-fit values of the BJMF model parameters used in our analysis in the quiescent and flaring state of TXS 0506+0506 blazar derived from the Lepto-hadronic models given by Ref. \cite{acciari2022investigating}. 

\begin{figure*}
    \centering
    \includegraphics[width=\textwidth]{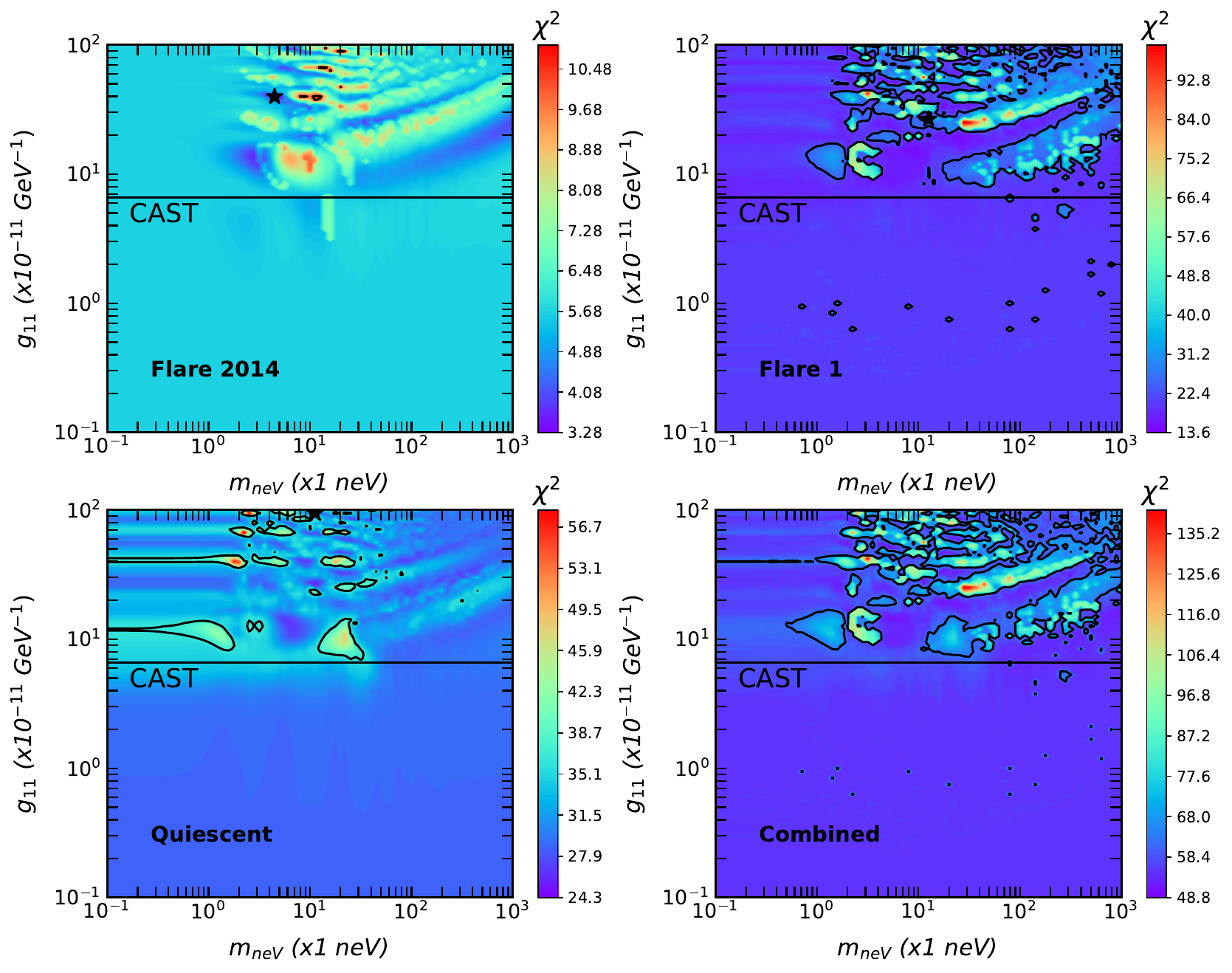}
    \caption{\label{fig:chi2dist}Distribution of $\chi^{2}_{ALP}$ in the $m_{neV}$-$g_{11}$ parameter space for all three phases. The $\star$ symbol in black represents the best-fit parameter point. The black contours represent the excluded parameter space at 95$\%$ C.L. in all three and the combined phases. The black horizontal line represents the upper limit set by the CAST experiment of g$_{a\gamma} <$ 6.6$\times$10$^{-11}$ GeV$^{-1}$ \cite{2017NatPh..13..584A}.}
\end{figure*}

Figure \ref{fig:chi2dist} shows the distribution of $\chi^{2}_{ALP}$ in the $m_{neV}-g_{11}$ parameter space for all three VHE activity phases. In Table \ref{table:bestfitsum}, we summarize the best-fit $\chi^{2}_{w/o ALP}$ and $\chi^{2}_{ALP}$ values along with the $m_{neV}-g_{11}$ parameters obtained under the null and ALP hypotheses. To summarize, we show the $\gamma$-ray flux for the best-fit $\chi^{2}$ values along with the final photon survival probability in Fig. \ref{fig:sedplots}.
\begin{table}
\begin{ruledtabular}
\caption{\label{table:bestfitsum}Summary of the best-fit $\chi^{2}$ values and ALP parameters under the null and ALP effect hypothesis for all phases.}
\begin{tabular}{c c c c c c}
 Phase & \textbf{$\chi^{2}_{w/o ALP}$} & \textbf{$\chi^{2}_{ALP}$} & $m_{neV}$ & $g_{11}$ & $\Delta \chi^{2}$   \\ 
 \hline
 Neutrino Flare 2014 & 5.65 & 3.31 & 4.47 & 39.81 & 6.72\\
VHE Flare 1 & 20.48 & 13.73 & 17.78 & 35.48 & 9.18\\
VHE Quiescent & 28.79 & 24.47 & 11.22 & 94.41 & 11.81\\
\end{tabular}
\end{ruledtabular}
\end{table}

To set the constraint on the $m_{a}$ -- $g_{a\gamma}$ parameter space, a threshold value $\chi^{2}_{thr} = \chi^{2}_{min} + \Delta\chi^{2}$ is determined to exclude the region at a certain C.L. for each phase. We generate 400 sets of pseudodata realized by Gaussian samplings as in Ref. \cite{liang2019constraints}, with the mean value and standard deviation taken as the best-fit flux under the null hypothesis and errors on the experimental data, respectively. 

\begin{figure*}[ht!]
    \centering
    \includegraphics[width=\textwidth]{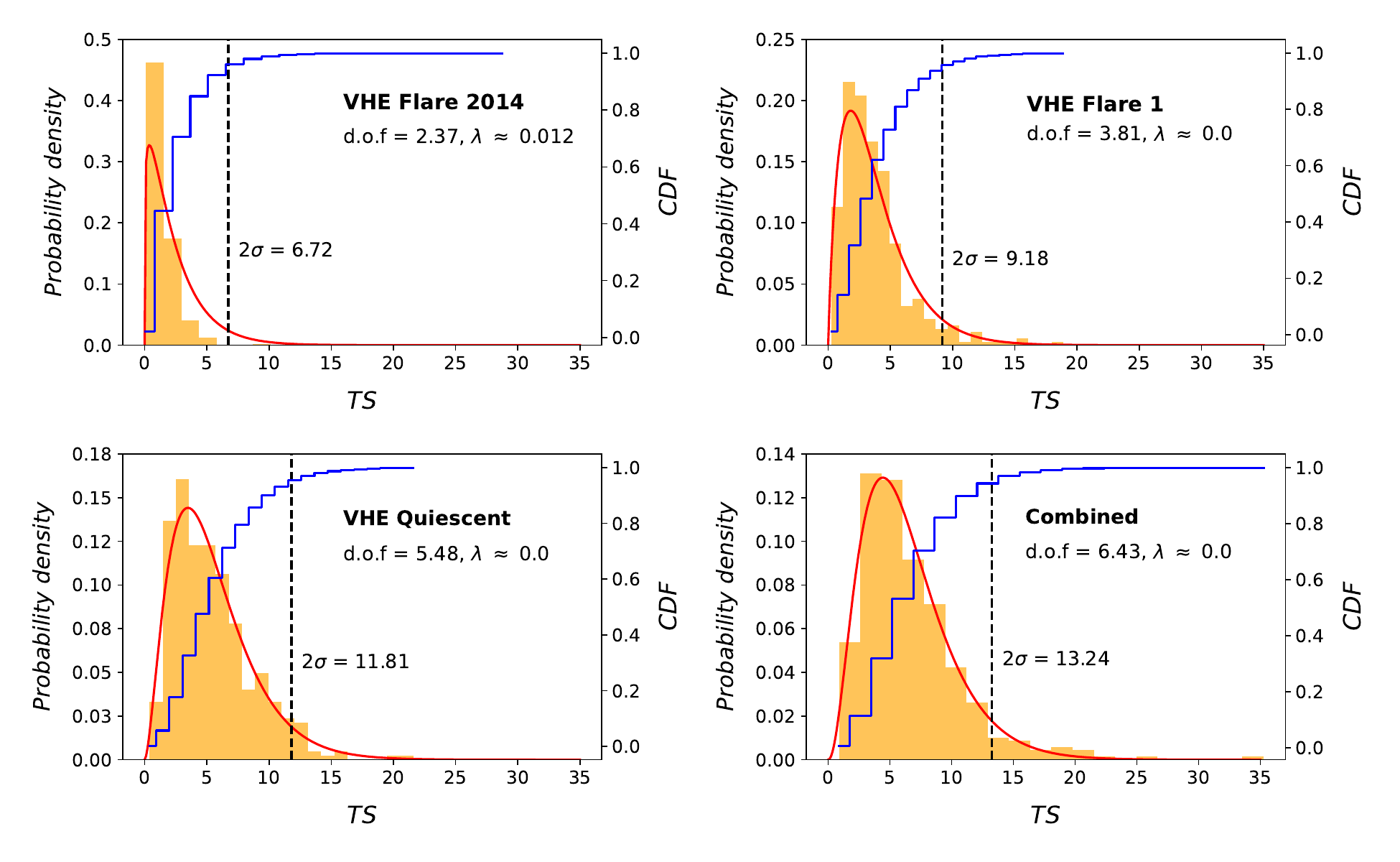}
    \caption{\label{fig:tsall}TS distributions of VHE Flare 2014 (top left), VHE Flare 1 (top right), VHE Quiescent (bottom left), and the combined (bottom right) phases of TXS 0506+056. The red lines show the fitted noncentral $\chi^{2}$ distributions. The blue lines show the cumulative density function (CDF) of the TS distributions.}
\end{figure*}

For each set, we calculate the best-fit $\chi^{2}$ for both the null and ALP hypotheses using the method described in Sec. \ref{sec:method}. We obtain the distribution of test statistics (TS) values,
%\begin{equation}
 $   TS = \chi^{2}_{null} - \chi^{2}_{ALP},
$%\end{equation}
 under the null hypothesis that obeys the noncentral $\chi^{2}$ distribution. The assumption that the probability distribution for the ALP scenario can be approximated with the null hypothesis, $\Delta \chi^{2}$, is derived at 95$\%$ C.L. as shown in Fig. \ref{fig:tsall}. The black contours in Fig. \ref{fig:chi2dist} represent the excluded parameter space at 95$\%$ C.L. We find the best constraint on $m_{a}$ -- $g_{a\gamma}$ parameter space in the Flare 1 phase. 

In this phase, we find a possible excluding constraint on $g_{a\gamma}$ that can go as low as $5 \times 10^{-11}$ GeV$^{-1}$ within 95\% C.L.. TXS 0506+056 is a variable blazar with a variability index of 245.9099, and significant determination of the intrinsic flux is difficult for a flaring blazar. Thus we obtain weaker constraints for ALP parameters. We even see this result in the combined analysis of the three phases.  On the other hand, for the Quiescent phase, this value could reach the similar constraint to that found by the CAST experiment.  
%{\color{blue} (IS IT FINE HERE OR PUT IT SOMEWHERE ELSE?) \color{red}The black contours in Fig. \ref{fig:chi2dist} represent the excluded parameter space at 95$\%$ C.L. in all three and the combined phases.}
\subsubsection{\label{subsec:resultsA2} Neutrino flare (2014-2015)}
We also investigate the ALP effect for the neutrino flare phase of the blazar TXS 0506+056. For the neutrino flare, there were no VHE observations. Hence we calculated the best-fit oscillation probability using only the \textit{Fermi}-LAT events following the method as in Sec. \ref{sec:method}. However, this phase could not add any significant result to the  $m_{a}$ -- $g_{a\gamma}$ parameter space. 

\subsection{\label{sec:resultsB}ALP effect at sub-PeV energies }
Extending this survival probability to sub-PeV energies, we can estimate the residual photon flux of the observed neutrino's counterpart. For this calculation, we use the ALP-$\gamma$ oscillation using the CAST upper limits on the parameters, namely $m_{a} = 1$ neV and $g_{a\gamma} = 6.6 \times 10^{-11}$ GeV$^{-1}$.

Assuming that IC170922-A resulted from proton-proton interactions, as calculated in Ref. \cite{PhysRevD.99.103006}, the average counterpart $\gamma$ rays will follow \cite{PhysRevD.78.034013},
%\begin{equation}
 $  E^{2}_{\gamma}\cdot \frac{dN_{\gamma}}{dE_{\gamma}} = \frac{2}{3} E^{2}_{\nu} \cdot \frac{dN_{\nu}}{dE_{\nu}}\, , \label{eq:nugamspect}
$%\end{equation}
%the observed astrophysical neutrino upper limit given by the IceCube experiment is generated entirely by the p-p interaction in the blob of the blazar jet. The associated $\gamma$-ray spectrum, which arises due to $\pi^{0}\rightarrow 2 \gamma$ decay, can be related to the neutrino spectrum via \cite{xyz}, 
 where $E_{\gamma}\approx 2E_{\nu}$ is the energy of the photons produced from $\pi^{0}$ decay. Subsequently, these VHE photons attenuate by interacting with the synchrotron and synchrotron self-Compton photons, resulting from relativistic electrons inside the blob. The fraction of VHE photons that can escape from the blob is estimated as in Ref. \cite{2013ApJ...768...54B},
 \begin{equation}
  \mathcal{F}^{esc}_{\gamma\gamma} = \frac{1-\exp{(-\tau_{\gamma\gamma}(\epsilon^{\prime}_{\gamma})})}{\tau_{\gamma\gamma}} \, , \label{eq:escfrac}
\end{equation}
%While propagating inside the jet, these VHE photons interacts with the isotropic distribution of low energy photons, which arises due to  in the comoving frame, and undergoes cascade. 
where the optical depth due to the interaction of high-energy photons of energy $\epsilon_{\gamma}'$ in the comoving frame is 

 \begin{equation}
  \tau_{\gamma\gamma}(\epsilon^{\prime}_{\gamma}) = R^{\prime}_{blob} \int_{\epsilon_{thr}} \sigma_{\gamma\gamma}(\epsilon^{\prime}_{\gamma}, \epsilon^{\prime}_{k}) \, n^{\prime}_{k}(\epsilon^{\prime}_{k}) \, d\epsilon^{\prime}_{k} \, .
\end{equation}
$n^{\prime}_{k}(\epsilon^{\prime}_{k})$ is the number density of the ambient photons of energy $\epsilon^{\prime}_{k}$ (in $m_{e}c^{2}$) in the comoving frame,
 \begin{equation}
    n^{\prime}_{k}(\epsilon^{\prime}_{k}) = \frac{2 D_{L}^2}{c R^{\prime2}_{blob}\delta^{2}\Gamma^{2}_{k}} \, \frac{F_{k}(\epsilon_{k})}{m_{e}c^{2}\epsilon^{\prime2}_{k}} \, ,
\end{equation}
and $\sigma_{\gamma\gamma}$ is the pair-production cross section \cite{10.1111/j.1365-2966.2008.13315.x}. $R^{'}_{blob}$ is the radius of the blob, and $D^{2}_{L}$ is the luminosity distance of the source. $\Gamma_{k}$ is the bulk Lorentz factor and $F_{k}(\epsilon_{k})$ is the photon flux in the observer frame.

\begin{figure*}[ht!]
    \centering
    \includegraphics[width=\textwidth]{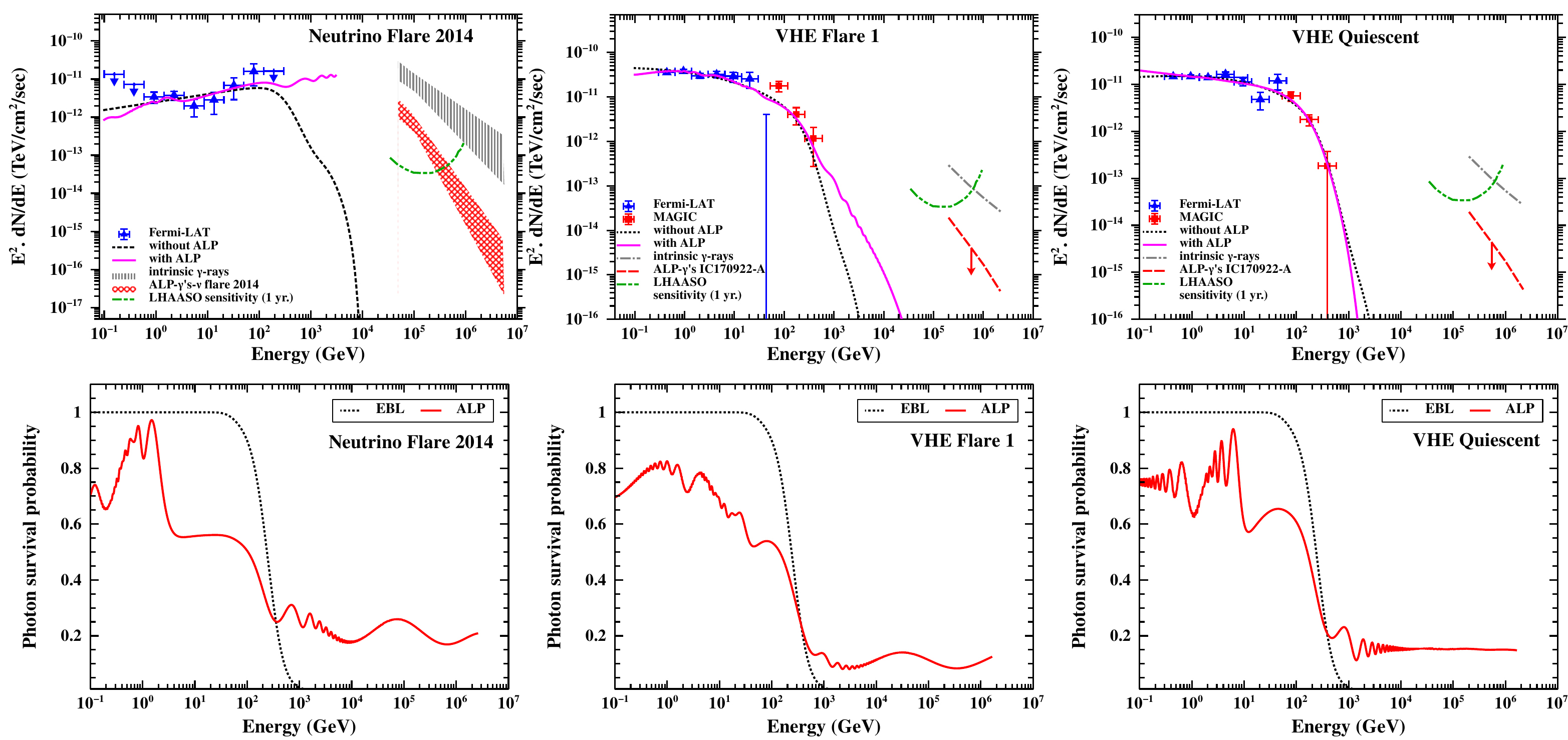}
    %\includegraphics[width=\textwidth]{SED_IC_TBT.pdf}
    %\textcolor{red}{(1) Above one or below? \\ (2) Is it a good idea to put CAST $\gamma$-rays? The ProbALP is not changing at all! Although we have different axion masses, the gamma flux from CAST is coming out the same for each phase.}
    %\includegraphics[width=\textwidth]{SED_wCAST.pdf}
    \caption{\label{fig:sedplots}Top: best-fit $\gamma$-ray spectra of TXS 0506+056 for all three VHE phases. The dotted black and the solid magenta curves represent the spectra under the null and ALP hypotheses, respectively. The corresponding best-fit $\chi^{2}$ values are listed in Table \ref{table:bestfitsum}. The experimental data are from \textit{Fermi}-LAT \footnote{see footnote \ref{foot:fermi}} and MAGIC \cite{acciari2022investigating}. The dot-dashed grey curves represent the intrinsic $\gamma$-rays and the dashed red curves represent the flux expected by considering the ALP effect (see text). The differential sensitivity to
a Crab-like point gamma-ray source for 1 year of exposure  by LHAASO is shown by the dash-dot-dotted curve (green) for comparison \cite{Vernetto_2016}. Bottom: final photon survival probability for best-fit $\chi^{2}_{ALP}$ parameter values under the ALP hypothesis.}
\end{figure*}
Hence, the intrinsic $\gamma$-ray spectrum can be obtained by counterpart gamma rays as in Sec. \ref{sec:resultsB} and this is shown by the grey vertical hatched and dot-dashed curves in the top panel of Fig. \ref{fig:sedplots}. This flux generally gets exhausted due to interaction with the EBL and CMB. On multiplying the intrinsic flux with the photon survival probability under photon-ALP oscillations for the corresponding phase, we can have a surviving fraction of the flux. The $\gamma$ rays due to the ALP effect at sub-PeV energies is shown by the red cross hatched and dashed curves. %\ref{fig:sedplots}.The survived $\gamma$-rays are shown , while the intrinsic photon flux is shown with the vertical-hatched lines in figure  
Interestingly, the surviving photons, considering the ALP-$\gamma$ oscillations in VHE Flare 2014, are above the differential sensitivity to a Crab-like point gamma-ray source for 1 year of exposure by LHAASO \cite{Vernetto_2016}..
%We also show the  with dash-dot-dot line (green) for comparison \cite{Vernetto_2016}.   
\subsection{\label{sec:resultsC}Diffuse $\gamma$ rays from FSRQs sources at sub-PeV energies due to ALP effect}
In this section, we calculate the diffuse flux at sub-PeV energies as an effect of ALP-$\gamma$ oscillation in sources like TXS 0506+056. The authors of Refs. \cite{10.1093/mnrasl/slz011, 10.1093/mnras/sty1852} emphasized that TXS 0506+056 cannot be considered a blazar of BL Lac type, but rather an intrinsically FSRQ, and all FSRQs are of the low-energy (synchrotron) peaked (LBL). Hence, we calculate the diffuse ALP $\gamma$ flux for the source luminosity function (LF) evolution like FSRQs using 
\begin{eqnarray}
    \Phi_{diff}(E_{\gamma}) &=& \int_{\Gamma_{min}}^{\Gamma_{max}} {\frac{dN}{d\Gamma} \, d\Gamma }\int_{z_{min}}^{z_{max}} \frac{d^{2}V}{dz d\Omega} \, dz \int_{L_{\gamma}^{min}}^{L_{\gamma}^{max}} \, dL_{\gamma} \, \nonumber \\ & &\times \,  \rho(L_{\gamma},z) .\, \frac{dF_{\gamma}^{int}}{dE} .\, e^{-\tau_{\gamma\gamma}^{alp}(E,z)} \, ,
\end{eqnarray}
where $dN/d\Gamma$ is the intrinsic photon index distribution which is assumed to be a Gaussian, $d^{2}V/dz d\Omega$ is the comoving volume element per unit redshift per unit solid angle, $dF_{\gamma}^{int}/dE$ is the intrinsic photon flux, here taken as calculated in Sec. \ref{sec:resultsB} for TXS 0506+056, $\tau^{alp}_{\gamma\gamma}$ is the opacity under the ALP scenario considering the CAST upper limit on the parameters  $m_{a}$ -- $g_{a\gamma}$, and $\rho(L_{\gamma},z)$ is the gamma-ray luminosity function (GLF).

We consider here the luminosity-dependent density evolution of the GLF with parametrization given by:
\begin{eqnarray}
   \rho(L_{\gamma},z) &=& \frac{A}{log(10).L_{\gamma}}\left[\left(\frac{L_{\gamma}}{L_{c}}\right)^{\delta1} + \left(\frac{L_{\gamma}}{L_{c}}\right)^{\delta2} \right]^{-1} \nonumber \\ & & \times \, \zeta(L_{\gamma},z) \,,
\end{eqnarray}
with
\begin{equation}
   \zeta(L_{\gamma},z) = \left[\left(\frac{1+z}{1+z_{c}(L_{\gamma})} \right)^{\eta1} + \left(\frac{1+z}{1+z_{c}(L_{\gamma})} \right)^{\eta2} \right] \, ,
\end{equation}
where
%\begin{equation}
   $z_{c}(L_{\gamma}) = z_{c}^{*}\left(\frac{L_{\gamma}}{10^{48}} \right)^{\alpha} \,$.
%\end{equation}
We collect the best-fit parameters $A, L_c, z_c^{*}, \alpha, \eta1, \eta2, \delta_1$, and $\delta_2$ for three blazar classes, namely, FSRQ \cite{2013zeng}, HSP, and LISP sources \cite{DiMauro_2014}. The limits of integration are $\Gamma_{min}=1.2$, $\Gamma_{max}=3.0$, $z_{min}>0$, and $z_{max}=2$. For FSRQs, we also calculate the diffuse flux using the integration limits of Ref. \cite{2013zeng} as a comparison.   %For the limits of $\L_{\gamma}$, we have considered two cases: (i) $\L_{\gamma} \in [10^{40}, 10^{52}]$ erg/sec, and (ii) $\L_{\gamma} \in [10^{42}, 10^{52}]$ erg/sec.

\begin{figure*}[ht!]
    \centering
    \includegraphics[width=0.7\textwidth]{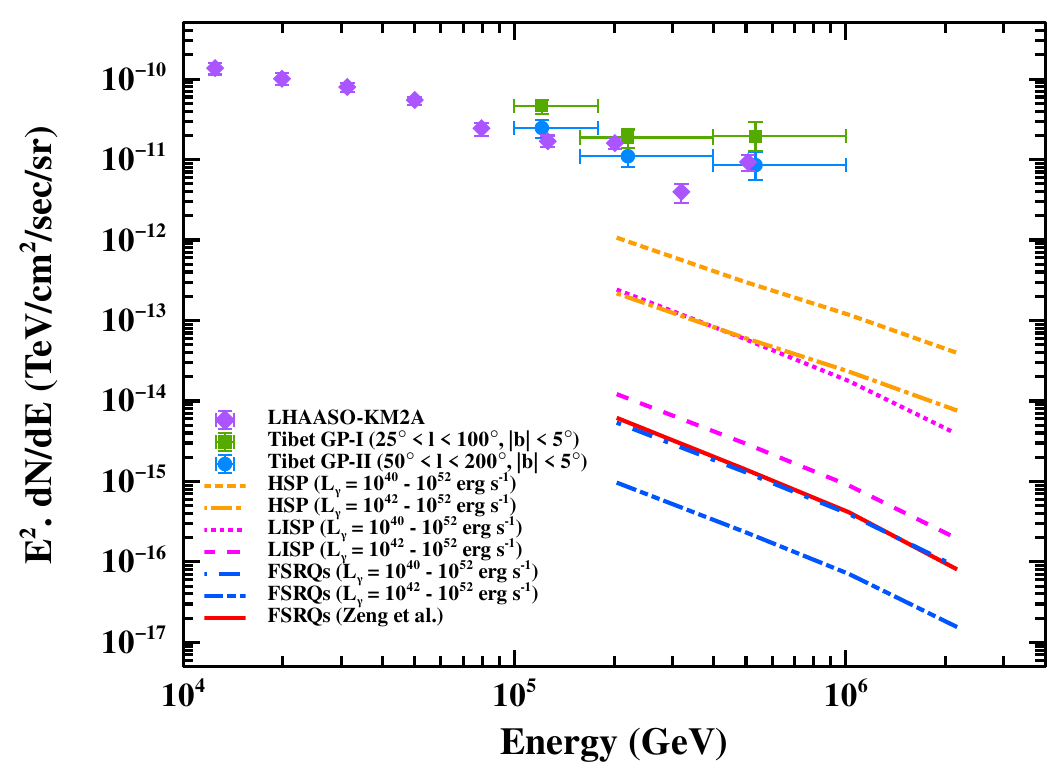}
    \caption{\label{fig:diffgalps} Expected diffuse $\gamma$-ray flux from photon-ALP conversion for FSRQ, HSP, and LISP sources. The data points are the Galactic diffuse gamma-ray emission measured by LHAASO-KM2A \cite{2022icrc.confE.859Z} (light purple) and Tibet AS-$\gamma$ \cite{2021amenomori} (blue and green). For comparison, the diffuse flux obtained using the parameters of Ref. \cite{2013zeng} is shown by the red solid curve.}
\end{figure*}

Figure \ref{fig:diffgalps} shows the diffuse $\gamma$ ray flux expected from photon-ALP conversion for the three classes. The data points of diffuse gamma-ray emission from the Galactic plane recorded by the LHAASO-KM2A \cite{2022icrc.confE.859Z} and Tibet AS-$\gamma$ \cite{2021amenomori} experiments are also shown. We show each class with (i) $\L_{\gamma} \in [10^{40}, 10^{52}]$ erg/sec, and (ii) $\L_{\gamma} \in [10^{42}, 10^{52}]$ erg/sec.
Dedicated surveys of the extragalactic diffuse gamma-ray flux by observatories like LHAASO, Tibet AS-$\gamma$, and the upcoming Cerenkov Telescope Array \cite{acharyacta} will be able to constrain the ALP parameters considering photon-ALP oscillations at sub-PeV energies as proposed here.  

\begin{acknowledgments}
The authors thank the anonymous referee for constructive comments which helped in improving the manuscript. The authors acknowledge the Science and Engineering Research Board (SERB) Grant No. SRG/2020/001932.
\end{acknowledgments}

\bibliography{references}

\end{document}